# ISiCLE: A molecular collision cross section calculation pipeline for establishing large in silico reference libraries for compound identification

Sean M. Colby[1], Dennis G. Thomas[1], Jamie R. Nunez[1], Douglas J. Baxter[1], Kurt R. Glaesemann[2], Joseph M. Brown[1], Meg A Pirrung[3], Niranjan Govind[1], Justin G. Teeguarden[1,4], Thomas O. Metz[1,*], Ryan S. Renslow[1,*]

[1] Earth and Biological Sciences Directorate, Pacific Northwest National Laboratory, Richland, WA, USA.

[2] Communications and Information Technology Directorate, Pacific Northwest National Laboratory, Richland, WA, USA.

[3] National Security Directorate, Pacific Northwest National Laboratory, Richland, WA, USA.

[4] Department of Environmental and Molecular Toxicology, Oregon State University, Corvallis, OR, USA

**ABSTRACT:** High throughput, comprehensive, and confident identifications of metabolites and other chemicals in biological and environmental samples will revolutionize our understanding of the role these chemically diverse molecules play in biological systems. Despite recent technological advances, metabolomics studies still result in the detection of a disproportionate number of features than cannot be confidently assigned to a chemical structure. This inadequacy is driven by the single most significant limitation in metabolomics: the reliance on reference libraries constructed by analysis of authentic reference chemicals with limited commercial availability. To this end, we have developed the *in silico* chemical library engine (ISiCLE), a high-performance computing-friendly cheminformatics workflow for generating libraries of chemical properties. In the instantiation described here, we predict probable three-dimensional molecular conformers (i.e. conformational isomers) using chemical identifiers as input, from which collision cross sections (CCS) are derived. The approach employs state-of-the-art first-principles simulation, distinguished by use of molecular dynamics, quantum chemistry, and ion mobility calculations to generate structures and libraries, all without training data. Importantly, optimization of ISiCLE included a refactoring of the popular MOBCAL code for trajectory-based mobility calculations, improving its computational efficiency by over two orders of magnitude. Calculated CCS values were validated against 1,983 experimentally-measured CCS values and compared to previously reported CCS calculation approaches. Average calculated CCS error for the validation set is 3.2% using standard parameters─outperforming other density functional theory (DFT)-based methods in the literature and MetCCS. An online database is introduced for sharing both calculated and experimental CCS values (metabolomics.pnnl.gov), initially including a CCS library with over 1 million entries. Finally, three successful applications of molecule characterization using calculated CCS are described, including the identification of an environmental degradation product, the separation and identification of molecular isomers, and the decoding of complex blinded mixtures of exposure chemicals. This work represents a promising method to address the limitations of small molecule identification, and offers alternatives to standards-based chemical identification amenable to high-performance computation.

The capability to routinely measure and identify even a modest fraction of biologically and environmentally important small molecules within all of chemical space—greater than $10^{60}$ potential compounds[5]—remains one of the grand challenges in biology and environmental monitoring. This long-term challenge is best met by analytical approaches capable of measuring broad classes of molecular species, referred to here as untargeted metabolomics. The technologies and driving concepts behind metabolomics have existed for nearly 40 years and have their origins in early metabolic profiling[6-12] and metabolic flux studies[13-14], as well as detection of metabolic defects and diagnosis of associated in-born errors of metabolism[15-17]. However, despite the solid foundation and the great strides made in metabolomics approaches over the past 20 years, present capabilities still fall short of comprehensive and unambiguous chemical identification of detected metabolites.

For example, NMR-based structural elucidation is an established method for unambiguous chemical structure assignment of novel molecules, but requires high sample concentration and purity. This limits its usefulness for high throughput and comprehensive structural elucidation. Synthesis of chemical reference standards for suspected novel molecules, is another alternative, but is costly, often difficult, and time consuming.

For identification of known molecules, the analytical methodologies that have proven to be the most efficient in confident identification of large numbers of metabolites in high throughput metabolomics studies have been GC-MS, LC-MS, and NMR, and comparison of experimental data to reference libraries containing data from analyses of authentic chemical standards using identical analytical methods. Such approaches adhere to the recommendations of the Metabolomics Standards Initiative of the Metabolomics Society for confident molecular identification[18-19], but depend on data from analysis of pure compounds. This represents a significant limitation, because authentic chemical standards are not available for the majority of metabolites[20]. For example, approximately 92% of the HMDB 4.0 molecules do not have authentic chemical standards (verified through custom Python scripts to search known vendors), and the HMDB only represents <5% of the estimated total metabolite space across multiple organisms[21-22]. Further, ChemSpider[23], PubChem[24], and American Chemical Society's chemical abstracts service (CAS) databases[25] contain entries for tens-of-millions of chemicals, yet one of the largest repositories of authentic reference spectra, the Wiley Registry / NIST Mass Spectral Library[26], contains data for roughly 730,000 unique compounds, <1% of known chemicals[27].

The most practical approach for dramatically increasing the size of reference libraries is through *in silico* calculation of molecular attributes. The metabolomics community has made great strides in predictions of chromatographic retention times and tandem mass spectra[28-32]. While the associated tools and methods have demonstrated important proofs-of-concept, challenges remain with relying on these predicted attributes for metabolite identification. For example, GC and LC separations involve interactions of molecules with surfaces, and degradation of chromatographic stationary phases will result in mismatch of experimental to predicted retention times. Tandem mass (MS/MS) spectra are gas-phase molecular properties, and less susceptible to the chemical interactions and artifacts that can affect retention time stabilities in GC and LC, and have good reproducibility between laboratories. MS/MS spectra for small molecules can be predicted with reasonable accuracy given appropriate training data[32-36], enabling the generation of short lists of molecules whose MS/MS spectra might match to experimental spectra. However, most MS/MS prediction methods rely on machine learning or deep learning approaches, and therefore can be limited by the size of the training data sets[36-37]. MS/MS spectra for molecules that are not chemically similar to compounds used in the training set may not be accurately predicted. New methods to accurately predict molecular properties that are also measured with high experimental reproducibility and without loss of data quality through time are required to transition metabolomics from the current paradigm to one applicable to global comprehensive chemical identification.

Quantum chemistry, i.e. the application of quantum mechanics to the understanding of molecules, holds great promise for calculation of molecular properties in support of global chemical identification. For example, infrared spectra[38], nuclear magnetic resonance chemical shifts[39], and molecular collisional cross sections (CCS) can be calculated from first principles, showing success where machine learning approaches[40] have underperformed. CCS is a measurable, calculable property of three dimensional chemical structures that can contribute to the unambiguous identification of even positional and *cis/trans* isomers[41-42]. CCS is a measure of the apparent surface area of a chemical ion and is related to the molecular gas-phase 3D conformation of that ion. It is reported as an area in angstroms ($Å^2$). Ion mobility (IM) spectrometry separates ions based on the extent of their interactions with an inert gas (usually $N_2$ or He) as they travel under the influence of an electric field[43]. Ions of smaller CCS have shorter drift times, while ions of larger CCS have longer drift times. CCS is highly sensitive to molecular shape, showing measurable differences between even positional and *cis/trans* isomers [41]. Ultra-high resolution IM separations coupled with mass spectrometry (IM-MS), such as structures for lossless ion manipulations (SLIM)[44-49], are capable of resolving compounds with only slight differences in stereochemistry and measuring mass and CCS with high accuracy. The gas-phase separations made by IM instruments have several advantages over conventional GC and LC platforms. Key amongst these are: extremely high reproducibility between instruments and laboratories (0.2% relative standard deviation)[50], no column degradation over time, separation principles sufficient to resolve constitutional, positional, and *cis/trans* isomers[41], and platforms currently advancing to provide separation resolution 5-fold higher than conventional platforms[44-49].

Currently, *in silico* methods for property prediction, including CCS, are limited by throughput, accuracy, and/or a reliance on large training sets. These obstacles hinder the rapid expansion of *in silico* libraries accurate enough for standards-free compound identification, particularly for molecules outside of any known training set (i.e. "out of sample"). To advance standards-free approaches through *in silico* methods, we introduce the *in silico* chemical library engine (ISiCLE), a quantum chemistry-based computational infrastructure for predicting molecular properties, including NMR chemical shift, infrared spectra, and CCS. The code is available upon request. We describe the development, optimization, and validation of the CCS calculation module of ISiCLE. We have architected ISiCLE for use on supercomputing resources, including a refactoring of the popular MOBCAL[51-53] code for trajectory-based mobility calculations, and validated the calculation of 1,983 CCS values against experimental data. Calculation accuracy is compared to similar first-principles approaches[54-55], as well as the property-based machine learning tool, MetCCS[40]. Finally, we provide a growing database of calculated CCS values, available at metabolomics.pnnl.gov, and a demonstration of the utility of calculated CCS in three examples.

## ■ MATERIALS AND METHODS

**Validation Set Molecules.** Lists of molecule standards and their measured CCS were collected from in-house data and from the literature[56-62]. Values were tabulated (see Supplementary Information file, SI_valset.xlsx) along with their associated CCS relative standard deviation, observed mass, IUPAC International Chemical Identifiers (InChI)[63], SMILES string, formula, chemical name, source citation DOI, and chemical class information. For details on how InChI were obtained and processed, please see the Supporting Information, Methods section. Only CCS collected on drift tube IM (DTIM) instruments with nitrogen ($N_2$) buffer gas were included, and only protonated, $[M+H]^+$, deprotonated, $[M-H]^-$, and sodiated $[M+Na]^+$ molecules were considered. If the CCS of the same compound and adduct ion (herein simply referred to as "adduct") were measured by two different sources, their CCS were included as two separate entries. All CCS values that were obtained in-house were collected using an Agilent 6560 Ion Mobility Q-TOF MS (Agilent Technologies, Santa Clara) with seven stepped electric field voltages, as described Zheng et al.[64].

Chemical class composition of the validation set, as determined by ClassyFire[2], is shown in Figure 1.

**ISiCLE CCS Calculation Module.** CCS values were calculated using ISiCLE, a high-throughput, automated computational pipeline built using the Python Snakemake framework[65], enabling scalability, portability, provenance, fault tolerance, and automatic job restarting. Snakemake is a workflow management system that provides a readable Python-based workflow definition language and execution environment that scales, without modification, from single-core workstations to compute clusters through as-available job queuing based on a task dependency graph. In addition, Snakemake workflows are human-readable and Pythonic. See the SI Discussion for a more thorough discussion on the benefits and justification for using a workflow engine. ISiCLE offers three calculation methods: *Lite*, *Standard*, and *AIMD* (*ab initio* molecular dynamics), each increasing in calculation accuracy and computational complexity. This work focuses on the *Standard* method, though *Lite* and *AIMD* methods are introduced and discussed. ISiCLE source code is available in the SI, and up-to-date versions are available for download on GitHub (github.com/pnnl/isicle).

The ISiCLE module for calculating IM CCS starts with the generation of 3D structures of ionized compounds (in .mol and .mol2 file formats) from a chemical structure identifier, such as the InChI of neutral parent compounds, and ends with the calculation of CCS values for various conformers of the ionized compounds using the trajectory method[51]. A conformer is any unique 3-dimensional arrangement of atoms for a molecule with the same bonding connectivity, i.e., a conformer is one of many constitutional stereoisomers of a molecule. For this work, protonated, deprotonated, and sodiated forms were considered for each molecule, but ISiCLE can be used to process other adducts as well (e.g., $[M+K]^+$, $[M+2Na]^{2+}$)[42].

The *Standard* pipeline, depicted in Figure 2, involves a series of intermediate steps for conformer generation using molecular dynamics (MD) simulations and geometry optimization using quantum chemical calculations, via density functional theory (DFT), on PNNL supercomputing resources. Each step of the CCS calculation pipeline is executed using a series of Python and shell scripts developed in-house, all coordinated through the Snakemake workflow. The details of each step in the Snakemake workflow are described below.

**InChI to 3D Structure Creation.** Each processed (desalted, neutral, major tautomer) InChI is converted to a two-dimensional (2D) representation of the compound using OpenBabel[66-67]. Three-dimensional structures are then generated and subsequently optimized using the generalized amber force field (GAFF)[68] in OpenBabel. Ionized forms of the neutral structures were generated by identifying ionization sites in each parent 3D structure based on pKa values, which were automatically calculated inline using the ChemAxon command line tool, *cxcalc*[4]. The strongest acidic atom (lowest pKa) was assigned as the deprotonation site, and the strongest basic atom (highest pKb) was assigned as the protonation and sodiation sites. All ionized structures were saved in the .mol2 file format.

**Conformer Generation and Geometry Optimization.** During experimental analysis of authentic chemical reference standards, a continuous distribution of CCS values or even multiple CCS values can be observed for a single ionized molecule or complex. This necessitates the use of *in silico* conformer sampling methods to capture the CCS distribution, lest *in silico* predictions fail to achieve required levels of accuracy[69]. Ionized structures are used to seed conformer generation by *in vacuo* MD simulations, using SANDER (Simulated Annealing with NMR-Derived Energy Restraints) from AmberTools17[70]. The atom partial charges are assigned using the AM1 bond charge correction (BCC) set[71-72] or determined from DFT calculations using NWChem 6.6[73]. The conformers are then generated in three stages. In the first stage, SANDER is used to perform a short energy-minimization (500 iterations) to relax the initial structure and to remove any spurious atom contacts. In the second stage, a short 50 ps MD run (with 0.5 fs time step) is performed to heat the structure from 0 to 300 K, without non-bonded cutoffs. In the third stage, 10 simulated annealing cycles are performed, where each cycle continues for a total of 1600 ps of 1 fs MD steps, with the following temperature profile: heating from 300 to 600 K (0 – 300 ps), equilibration at 600 K (300 – 800 ps), cooling from 600 to 300 K (800 – 1100 ps), and equilibration at 300 K (1100 – 1600 ps). Ten conformers from the low-temperature equilibration stage (300 K) of each simulated annealing cycle were randomly selected and then down-selected to three by identifying the two most dissimilar conformers and the single most similar conformer, leading to a total of 30 conformers. The dissimilar conformers were determined as the two conformers with the largest pairwise root-mean-square deviation (RMSD) of their atomic positions, while the most similar conformer had the lowest pairwise RMSD sum among the 10 conformers. The three selected conformers were sufficiently representative of the ten conformers in a single simulated annealing step[41]. Thus, a total of 30 conformer geometries are used for subsequent geometry optimization with DFT using NWChem.

**Density Functional Theory Calculation.** To further optimize resulting molecular geometries, quantum chemical DFT calculations were performed using NWChem, an open-source, high-performance computational chemistry software developed at PNNL, similar to methods described in previous studies[42, 74]. The B3LYP exchange-correlation functional was used for all energy and geometry optimization calculations[75-78]. All basis sets were obtained from the Environmental Molecular Sciences Laboratory (EMSL) Basis Set Exchange[79-80], which included the Pople basis set at the 6-31+G** level (a double-zeta valence potential basis set having a single polarization function)[81-83].

**CCS Calculation via MOBCAL-SHM.** CCS values of the geometry-optimized conformers were calculated using the trajectory method, as implemented in our new version of MOBCAL, called MOBCAL-SHM (shared memory; see Results). MOBCAL-SHM source code and binaries are available in the SI, and up-to-date versions are available for download on GitHub (github.com/pnnl/mobcal-shm). MOBCAL[51-53] was selected among comparable alternatives as its implementation of the trajectory method is generally accepted to be the "gold standard" for computational CCS calculation[84-88]. The original version of MOBCAL is computationally intensive; therefore, to improve computational efficiency, we developed and optimized the parallel MOBCAL-SHM, written in C. Briefly, modifications to MOBCAL included (i) use of shared memory for parallel communication, (ii) use of dynamic load balancing instead of round robin, (iii) reorganization of the *dljpot* computations to remove replicated computations and enable use of vector registers, (iv) reduction of the number of cache lines being accessed per loop iteration by interleaving the data vectors, and (v) reordering of the *ig* and *ic* loops in *mobil2*. A shared memory implementation was chosen, as opposed to a message passing interface (MPI) implementation, in order to enable scaling on increasingly parallel single-node architectures, e.g. 68 cores per node on Intel's Knight's Landing coprocessors.

**Averaging Calculated CCS Values of Conformers for Comparison to Experimental Values.** Reported CCS values are normally a single value per adduct and often chosen based on experimental signal strength, centroid analysis, and relative CCS peak location (e.g., to avoid selecting the CCS of a multimer). Thus, to calculate a single CCS value for each ionized structure from a set of conformers, a number of methods were evaluated, including methods similar to those implemented by Paglia et al.[54] and Bowers et al[55]. Putative methods resulted from the Cartesian product of three sets: (i) optimization scheme, (ii) number/type of conformers used in the average, and (iii) averaging method.

*I. Optimization scheme.* The Optimization scheme explores whether the final geometry optimization by DFT is necessary to achieve lowest error, as it is the most computationally intensive step in the pipeline. Thus, we performed DFT calculations for each conformer to two levels of efficacy: optimization for energy only, and optimization for energy and structure.

*II. Number/type of conformers.* Methods of sampling conformers from the MD step, which produces two least-similar conformers and a single most-representative conformer for each of the 10 simulated annealing steps, were also evaluated. This set therefore includes use of all sampled conformers (30), only the least similar conformers (20), and the most representative conformers (10).

*III. Averaging method.* Averaging methods included the mean and median CCS of conformers for each ionized structure, as well as three energy-based methods: (a) CCS of the lowest energy conformer, (b) the mean CCS of those conformers with relative energy less than 5 kcal/mol, and (c) the sum of each conformer's CCS contribution, Boltzmann-weighted by relative energy. We hypothesized that Boltzmann weighting, based on calculated DFT energies, would shift the overall CCS distribution toward higher probability conformers, thus creating CCS distributions that are characteristic of IM experiments.

Combined, the two optimization schemes, three conformer sampling methods, and five averaging methods resulted in 30 putative approaches for reducing a distribution of CCS values across conformers to a single CCS value per ionized structure. A comparison of these approaches with respect to mean absolute error (MAE) is summarized in Table S1, which includes results of similar approaches. The method introduced by Paglia et al. is similar to taking the lowest-energy conformer among all sampled conformers, DFT optimized for energy only. The method introduced by Bowers et al. is similar to averaging all energy- and structure-optimized conformers with relative delta energy less than 5 kcal/mol.

**Calibration.** Additional steps were taken to account for inaccuracies in the various components of the pipeline. For example, significant departures from a plot of CCS vs mass signals an anomalous calculation, as mass and CCS have been shown to trend quite closely (correlation coefficient: 0.92)[89]. To this end, we constructed a fit of mass versus predicted CCS and filtered calculations that lie outside a 98% confidence interval of the regression (Figure S1).

Additionally, we calibrated calculations against the experimental standards by linear regression, fit to minimize mean absolute error (MAE). Calibrations were performed per-adduct, as increased error, on average, was observed for sodiated compounds in comparison to other adduct types (protonated, deprotonated). For pre-calibration results, see Supporting Information, Results and Discussion section.

**Lite Method.** For applications that do not require as high CCS accuracy, a *Lite* method was created for rapid calculation. Adducts were created in the same way as described for the *Standard* method, but instead of processing resulting adducts using MD and subsequent steps, CCS was calculated for each adduct directly using IMPACT[89] with the following settings: take hydrogens into account, 64 shots per rotation, 0.001 convergence threshold, and 64 independent runs. Resulting CCS values assume a helium buffer gas ($CCS_{He}$), which we converted to nitrogen buffer gas-based CCS ($CCS_{N2}$) using Equation 1.

$$CCS_{N2} = CCS_{He} + \alpha m^\beta \text{ (Eq. 1)}$$

Where m is the mass of the parent compound. Parameters α and β, determined using least squares minimization by fitting $CCS_{He}$ output from IMPACT to experimental $CCS_{N2}$ values, are 27.9 $z\text{Å}^2/m$ and 0.14 (dimensionless), respectively.

***Ab initio* Molecular Dynamics (*AIMD*)-based Method.** For applications that require higher CCS accuracy, at the cost of additional computational time, the *AIMD-based* method was created. Adducts were created in the same way as described for ISiCLE *Standard*. However, instead of performing MD and DFT steps separately (i.e. MD to generate a conformer distribution, DFT to optimize the resulting geometries), *ab initio* molecular dynamics (AIMD)[90], implemented in NWChem, was used to simultaneously generate and optimize (with respect to geometry and energy) conformers, which were ultimately averaged by Boltzmann weighting.

**Architecture.** ISiCLE was evaluated and implemented on PNNL supercomputing resources, *Constance* and *Cascade*. Each of the 464 nodes of the *Constance* supercomputer is dual socket, equipped with 12-core Intel Haswell processors E5-2670v3 (running at 2.3 GHz) for a total of 24 cores per node, and 64 GB of DDR3-1600 memory. Each of the 1,440 nodes of the *Cascade* supercomputer is equipped with 16 Intel Xeon cores (E5-2670, running at 2.6 GHz) and 128 GB of memory. Both computers' nodes are connected by a Fourteen Data Rate (FDR) InfiniBand fabric. Runtimes are reported in node-hours, as ISiCLE is designed to scale across arbitrary HPC resources.

After analysis by ClassyFire[2], our validation set was found to have 14 chemical superclasses and 76 classes. A total of 1,308 unique compounds are included with masses ranging from 68.0374 Da to 1072.3806 Da. Both the *Standard* and *Lite* methods of ISiCLE were used to generate calculated CCS values for the validation set.

***In Silico* Reference Library and Online Database.** In addition to the validation set, ISiCLE was used to calculate CCS values for the HMDB, Universal Natural Product Database (UNPD)[91], and the Distributed Structure-Searchable Toxicity (DSSTox) Database[92]. CCS values (protonated, deprotonated, and sodiated forms) were calculated using the ISiCLE *Lite* method for compounds from several databases that fell within the 50 to 1100 Da mass range (~80k from the HMDB, ~205k from the UNPD, and 720k from the DSSTox). Additionally, some compounds from the HMDB were run through ISiCLE

*Standard* CCS calculations. All of these values are available at metabolomics.pnnl.gov, which is being regularly updated to expand the number of compounds and to replace *Lite* CCS values with *Standard* CCS values as they become available.

## ■ RESULTS

The efforts detailed in this work produced ISiCLE to address long standing challenges hindering identification of the vast set of features in complex biological samples for which standards do not exist. Standards-free identification of small molecules requires accurate calculation of properties that can be reliably measured experimentally, such as CCS. *In silico* methods must also be fast enough to make scientific contributions on a meaningful time scale, especially when cultivating libraries of *in silico* properties large enough for robust and comprehensive compound identification. Moreover, methods must not rely entirely on reference standards or training sets, as these impose limitations on novel molecule identification and discovery. The following results demonstrate ISiCLE's success in terms of accuracy, achieving 3.2% unsigned error; throughput, processing molecules in a matter of hours; and out-of-sample generalization in cases where other approaches have failed.

**Mobility Calculation Improvements.** Improvements to MOBCAL were evaluated by comparing the average computation time of 10 representative ionized structures of various size. Timings are reported as factor speedup of MOBCAL-SHM over the original MOBCAL version on a per-node basis (Figure 3). CCS calculations made using MOBCAL-SHM differed from those of MOBCAL by 0.34%, attributable to differences in the pseudo-random number generators used: RANLUX[93] and Mersenne twister[94] for MOBCAL and MOBCAL-SHM, respectively.

It is worth noting that MOBCAL is a serial code; while MOBCAL-SHM is able to make use of all cores available on a single node, MOBCAL is limited to use of one. MOBCAL occupies the entire node during computation, and our efforts to instantiate multiple MOBCAL calculations on a node have been unsuccessful. Thus, per node, MOBCAL-SHM reduced average CCS computation time from 10.8 node-hours to 0.08 node-hours, amounting to a 143-fold increase in efficiency. For comparison, Zanotto and co-workers recently reported a 48-fold efficiency increase in a refactored version of MOBCAL[95]. MOBCAL-SHM is able to take advantage of on-node efficiencies at the expense of not generalizing to multi-node computation. For the validation set used in this work, computations were fast enough on a single node as to not require/benefit from use of multiple nodes and, by extension, an MPI version. Larger, more complex molecules, however, would incentivize adoption of a multi-node implementation.

**Computational Efficiency.** Each component of the ISiCLE CCS calculation pipeline is associated with a varying degree of computational demand, both in terms of number of operations and processing time required per operation. For example, InChI preprocessing is performed per parent molecule, 3D structure generation and MD calculations are performed per adduct, and DFT and mobility calculations are performed per conformer. Because comparisons to experimental data are made per adduct, we report the time required to process a single adduct end-to-end. Table 1 compares the computational burden of each step in the *Standard* pipeline, reported based on the average time required to process an adduct.

Altogether, the average time taken to process an adduct with ISiCLE *Standard* is 37.2 node-hours. Computation time heavily depends on the number of atoms that comprise a given structure, so total timings are further discretized by molecule size in Figure 4.

**Validation.** Among our explored approaches for averaging calculated CCS values of conformers (Table S1), the lowest error was similar for the top several methods. These included (i) DFT optimization of energy and structure, (ii) averaging over either 20 or 30 conformers, and (iii) averaging by taking the minimum-energy conformer or by Boltzmann weighting. Because Boltzmann weighting by energy offers theoretical improvements over minimum-energy methods[39], it was selected for the *Standard* method of ISiCLE. Additionally, following linear calibration, the Boltzmann method yields the lowest error. Figure 5 shows calculated CCS results for the validation set, plotted against *m/z*.

ISiCLE achieves 3.2% mean absolute error when evaluated against experimental CCS values. Compared to other methods of CCS calculations on the same set of molecules, ISiCLE performs significantly better. Methods developed by Paglia et al. and Bower et al. achieve errors of 5.3% and 5.2%, respectively. The MetCCS approach achieved a mean absolute error of 3.3%.

**Applications.** To demonstrate the utility of ISiCLE, we used calculated CCS, mass, and other properties in three example applications for the identification of small molecules in real samples.

*Application 1 - Degradation products in sediment*. Environmental samples of New York/New Jersey Waterway Sediment (NIST SRM 1944[96]) were analyzed by DTIM-MS, with determination of accurate mass and CCS features for multiple compounds. CCS was calculated *in silico* using the *Lite* method of ISiCLE for 21 possible degradation products (e.g. 2-hydroxyfluorene, 3-hydroxyflourene, 4,5-pyrenediol[97-99]) of 9 polycyclic aromatic hydrocarbons (e.g. fluorene, pyrene, 1,6-dimethylphenanthrene) present in the sediment. Parent compounds and predicted degradation products were identified by comparing only measured and calculated accurate mass and CCS. For example, experimental data for 4,5-pyrenediol[97-99] matched the predicted values within 1.1% (Table 2, representative data shown in Figure S2).

*Application 2 – U.S. Environmental Protection Agency (EPA) Non-Targeted Analysis Collaborative Trial (ENTACT) challenge*. We participated in the ENTACT inter-laboratory challenge[100], designed for the objective testing of non-targeted analytical chemistry methods using a consistent set of blinded synthetic mixtures. Each mixture contained an unknown number of chemicals (later revealed to be 95 to 365 compounds) in dimethylsulfoxide. All compounds were selected from the EPA ToxCast chemical library[101]. Further details on ENTACT are outlined in Sobus et al.[102] and Ulrich et al.[103] Calculated CCS was used in this study to identify compounds, along with high resolution mass and isotopic signature[100]. For CCS calculations, the ISiCLE *Standard* method was used for 16% of molecules in the ToxCast chemical library and the *Lite* method was used for the remaining molecules. In the end, our ToxCast CCS library had values for 11,633 adducts. Calculated CCS increased the confidence of 84% of molecules that were correctly determined to be present in the samples, showcasing its importance in this multi-attribute approach. Compared to the true positive experimental standards spiked in these samples that were

uniquely identified, calculated CCS errors for *Standard* and *Lite* methods of ISiCLE were 3.1% and 5.4%, respectively (Table 2). This out-of-sample test demonstrates consistent CCS error values compared to the initial validation set. Experimental and calculated CCS values from this study are available in the library introduced below.

*Application 3 – High accuracy CCS for positional and cis/trans isomers of chlorogenic acids*. We recently reported the ability of SLIM-MS to provide ultrahigh resolution IM separations[41] of positional and *cis/trans* isomers of dicaffeoylquinic acids (diCQAs), chlorogenic acids with reported anti-HIV and anti-inflammatory benefits[41]. Experimental CCS and CCS calculated using the ISiCLE *AIMD-based* method, were compared for 3,5-diCQA isomers. To further evaluate the accuracy of ISiCLE, we expanded the calculations to encompass all eight reported diCQA isomers, including 1-*trans*,3-*trans*; 1-*trans*,5-*trans*; 3-*trans*,4-*trans*; 3-*cis*,5-*cis*; 3-*cis*,5-*trans*; 3-*trans*,5-*cis*; 3-*trans*,5-*trans*; and 4-*trans*,5-*trans*-diCQA. Mean absolute error (MAE) results for the set were 4.8%, 2.6%, and 0.8% for *Lite*, *Standard*, and *AIMD-based* methods of ISiCLE (see Figure S3), respectively, compared to 6.4% for MetCCS. This out-of-sample set—i.e., set of compounds not present during model training—clearly demonstrates the performance-accuracy tradeoff and reveals sub-1% error when the *AIMD-based* method is used. As computational power increases, all CCS calculations in the near future could be performed with the *AIMD-based* method, with errors approaching, and potentially outperforming, those of experimental measures. This example also reveals one of the drawbacks of machine learning approaches that do not consider 2D or 3D molecular structures in their CCS calculation, such as MetCCS. The training parameters for these methods do not sufficiently differ between isomers to accurately distinguish their CCS values. Conformer consideration and 3D electron structure calculations alleviate this issue and can more accurately reflect the experimentally observed CCS values.

*In Silico* **Reference Library and Online Database.** CCS values for [M+H]$^+$, [M-H]$^-$, and [M+Na]$^+$ adducts are made available at metabolomics.pnnl.gov, currently totaling 1,455 and over 1 million entries for experimental and calculated values, respectively. This community resource will be updated as more values become available. The website provides additional information, including chemical name, SMILES, InChI, 2D structure, formula, and mass.

## ■ DISCUSSION

**Importance of Standards-Free Small Molecule Identification.** ISiCLE enables a departure from the reliance on experimentally derived chemical properties for complex mixture characterization. Determined by analysis of pure samples, experimental characterization of standards is an expensive, time-consuming practice that cannot accommodate candidate molecules that are (i) without a form available for purchase; (ii) without a protocol to synthesize; or (iii) as of yet undiscovered. ISiCLE enables expansion of chemical property libraries through calculation and, although initially dependent on experimental standards for calibration and validation, it will ultimately see use as a generative approach for creating significantly larger chemical property libraries than are currently possible.

Without reliance on experimental standards, characterization of complex samples becomes tractable with a sufficiently representative library. As ISiCLE evolves toward greater accuracy and number of calculated properties (CCS, NMR chemical shifts, and beyond) and more molecules are added to the *in silico* reference library, compound identification can be confidently made without experimental standards. It will be the work of organizations such as the Metabolomics Standards Initiative and metabolomics societies to establish frameworks and criteria for assessing the confidence of "identifications" made with *in silico* libraries. As an estimated >99% of metabolites are currently undiscovered[5, 104-106], accommodation of computational methods with confidence is imperative for the advancement of our fields.

**A Library for the Molecular Universe?** The nearly million-compound library reported here is a transformational increase over existing libraries. Nonetheless, as a minor fraction of chemical space, libraries of its size alone are not sufficient for identifying all reported features in complex biological samples. We recognize the likelihood that features will emerge from untargeted analyses that match no library entries because they represent a currently unknown chemical structure. For these cases, measured attributes such as high accuracy mass and isotopic signature can be used to generate plausible molecular structures, for example, through *in silico* metabolism simulators[107], deep learning-based neural networks[108], or with more effort, combinatorial searching of a given formula. Additional attributes of these new molecular structures—CCS, NMR chemical shifts, retention times, and MS-MS spectra—can then be calculated, and where sufficient data exist, be used to identify the subset mostly likely to represent the feature. Thus, ISiCLE can be used to generate attributes of probable chemical structures, with errors small enough to support down selection and provisional identification. A growing library of these new compounds would eventually come to represent an increasing portion of molecular space.

Rapid and extensive growth of *in silico* libraries is also an attractive approach to reduce the number of unidentifiable features in complex samples. Processing all molecules available in databases such as HMDB, UNPD, PubChem, and others[23, 25-26], using tools like ISiCLE, would establish libraries of our known or recorded molecular universe. This level of library expansion would eventually cover the majority of known biologically relevant molecules to include those for which authentic chemical standards are not available.

**Comparison with other CCS Calculation Methods.** Structure-based approaches that utilize first principles of quantum chemical calculations leverage our understanding of the underlying physics to predict chemical properties directly. Compared to approaches that predict chemical properties without first-principles simulation, such as the machine learning-based MetCCS, ISiCLE performs comparably with decreased CCS error, but with increased computation time. However, ISiCLE offers promise in that it will theoretically generalize more effectively to out-of-sample characterizations, a critical factor in growing a standards-free chemical property library. Machine learning methods, like MetCCS, are limited by the size and scope of the initial training set, and thus ultimately limited to the number of authentic chemical standards available for purchase. Furthermore, machine learning is challenged by chemicals with similar properties and similar structures, such as constitutional and configurational isomers (e.g., *cis/trans* isomers), as demonstrated above in Application 3 with diCQA. The input properties required for MetCCS were nearly identical for all 8 isomers, despite CCS values for this set spanning a range of over 43 Å$^2$, leading to predicted CCS errors as high as 9.5% (1-*trans*,5-*trans*-diCQA). We have demonstrated that our approach can surmount this challenge, and with high accuracy (average unsigned accuracy of 0.8% for this set). In addition, ISiCLE offers scalability across HPC resources, portability, provenance, and fault tolerance.

**ISiCLE Methods.** The ISiCLE module for calculating IM CCS for molecules has three methods for calculating CCS – *Lite*, *Standard*, and *AIMD-based* – of which the *Lite* and *Standard* methods were fully evaluated against the validation set of experimental values, and the *AIMD-based* method demonstrated in a specific application. Each method offers trade-offs between computational efficiency and calculation accuracy. The *Lite* method, with the lowest accuracy, requires the least amount of computational resources, and the *AIMD-based* method, with the highest accuracy to-date, requires significant computational resources. The *Standard* method offers a balance between accuracy and computational time, while still outperforming the accuracy of typical CCS calculation methods found in the literature. When considering a specific application, it is important to weigh the pros and cons of each implementation. With regard to the validation set used in this work, and the selected application, the forms of ISiCLE were selectively applied to each dataset depending on its size and the required level of accuracy.

The implementations of these different methods stress an important aspect of ISiCLE as a concept: the pipeline is modular such that as new algorithms and methods become available, they can replace analogous components of ISiCLE to improve computational efficiency, accuracy, and/or openness. Validation will still be necessary as components change, but the flexibility allows for a framework that can adapt to the state-of-the-science. With respect to improving openness, several components of ISiCLE currently require licenses and/or fees (*cxcalc*, ambertools) and will thus be the first targets for replacement to ensure ease of adoption.

**Computational Efficiency.** Our default method, the *Standard* method of ISiCLE, took an average of 37.2 node-hours per conformer. Compared to MetCCS this added computational cost yields a marginal improvement to calculation accuracy. While there are significant advantages to ISiCLE, as mentioned earlier, we look to improve accuracy, computational efficiency, or both, as a next step. When considering non-*Lite* methods of ISiCLE (*Standard*, *AIMD-based*), the most demanding computational steps are those involving DFT. For the *Standard* method, geometry optimization by DFT amounts to approximately 90% of the total computation time, the next highest being mobility calculations at 7%. Thus, to further improve computational efficiency without sacrificing accuracy, our focus in the future will shift from MOBCAL, the bottleneck when this work was started, to speeding up the DFT calculations with NWChem.

## CONCLUSION

In this manuscript we present the development of ISiCLE, a computational tool for accurate and supercomputing-enabled prediction of chemical properties using quantum chemical methods. This work offers 1) the first open-source, scalable (from desktop to HPC resources), and portable quantum chemistry-based collision cross section calculation workflow for the community, 2) an advanced conformer sampling method for higher accuracy property prediction, based on Boltzmann weighting to ensure that highly probable conformers are more represented, 3) a refactoring of the gold standard mobility calculation method (MOBCAL) with a speedup of over 2-orders of magnitude, 4) validation of the whole pipeline on the largest experimental dataset in the literature to date (unique values), 5) comparison of our approach with those in the literature, including competing machine learning approaches, and 6) a public library of over 1 million entries, covering the Human Metabolome Database, the EPA DSSTox exposure database, and the Universal Natural Product Database.

The transformation of the field of metabolomics towards global comprehensive identification of compounds in complex samples is underway. Among the many innovations that are necessary to reach this goal – e.g. ultrahigh resolution separation, higher throughput NMR – the development of *in silico* libraries of chemical attributes for identification of the multitude of compounds for which authentic standards don't exist, is a critical step. Development of ISiCLE, which included a two orders of magnitude improvement in the efficiency of MOBCAL, is an important first step towards meeting the goal of establishing large scale *in silico* reference libraries. ISiCLE has an easy to use software package for calculating chemical properties, including CCS, incentivizing adoption.

ISiCLE's *AIMD-based* method produced CCS values with absolute errors of 0.8%, approaching measurement error where the less computationally intensive implementations each had absolute errors less than current methods. Looking forward, ISiCLE's reliance on first principles and full 3D chemical structures may provide advantages over machine learning approaches derived from 2D structural information, particularly for positional isomers. Our successful use of ISiCLE for identification of the diCQA positional isomers highlights this import point for the field.

Recent funding of Compound Identification Development Cores by the National Institutes of Health reflects growing recognition of the challenge to the community that our limited libraries and availability of chemical standards pose. As momentum in the development of new methodologies for chemical identification through innovations in computational and experimental methods grows, a parallel need to consider best practices for use of these methods for chemical identification will also have to be fostered. In addition, it is clear to us that calculation of additional attributes – NMR chemical shifts, IR spectra, and others – will be necessary to increase the dimensionality of the array of attributes for chemical identification. Together, these improvements will help bring about the required paradigm shift away from reference-material based library building, and as a consequence, a rapid advancement in compound identification and biomedical discovery.

## ■ ASSOCIATED CONTENT

### Supporting Information

The Supporting Information is available free of charge on the ACS Publications website at DOI:

## ■ AUTHOR INFORMATION

### Corresponding Authors

\* Ryan S. Renslow - ryan.renslow@pnnl.gov
\* Thomas O. Metz - thomas.metz@pnnl.gov

## Author Contributions

The manuscript was written through contributions of all authors. All authors have given approval to the final version of the manuscript.

## ■ ACKNOWLEDGMENT

This research was supported by the Genomic Science Program (GSP), Office of Biological and Environmental Research (BER), U.S. Department of Energy (DOE), and is a contribution of the Metabolic and Spatial Interactions in Communities Scientific Focus Area. Additional support was provided by the Pacific Northwest National Laboratory (PNNL), Laboratory Directed Research and Development program via the Global Forensic Chemical Exposure Assessment for the Environmental Exposome project and the Microbiomes in Transition Initiative, and the National Institutes of Health, National Institute of Environmental Health Sciences grant U2CES030170 and National Cancer Institute grant R03CA222443. Ion mobility spectrometry-mass spectrometry analyses were performed in the Environmental Molecular Sciences Laboratory (EMSL), a national scientific user facility sponsored by BER and located at PNNL. Simulations were performed on PNNL Institutional Computing and EMSL high-performance computing resources. PNNL is operated for DOE by Battelle Memorial Institute under contract DE-AC05-76RL01830. The authors thank Dr. Neeraj Kumar (PNNL) for valuable discussions on the MD simulations, and we thank Dr. Erin Baker (NCSU) and Dr. Samuel Payne (BYU) for discussions regarding the experimental data and statistics.
## ■ ABBREVIATIONS

DNA, deoxyribonucleic acid; RNA, ribonucleic acid; NMR, nuclear magnetic resonance; MS, mass spectroscopy; GC, gas chromatography; LC, liquid chromatography; HMDB, human metabolome database; CAS, chemical abstracts service; CCS, collision cross section; 3D, three-dimensional; IM, ion mobility; SLIM, structures for lossless ion manipulations; ISiCLE, *in silico* chemical library engine; PNNL, Pacific Northwest National Laboratory; InChI, international chemical identifier; MD, molecular dynamics; DFT, density functional theory; 2D, two-dimensional; GAFF, generalized amber forcefield; RMSD, root-mean-square deviation; MPI, message-passing interface; MAE, mean absolute error; HPC, high-performance computing; AIMD, *ab initio* molecular dynamics; diCQA, dicaffeoylquinic acid.

## ■ REFERENCES

1. Feunang, Y. D.; Eisner, R.; Knox, C.; Chepelev, L.; Hastings, J.; Owen, G.; Fahy, E.; Steinbeck, C.; Subramanian, S.; Bolton, E.; Greiner, R.; Wishart, D. S., ClassyFire: automated chemical classification with a comprehensive, computable taxonomy. *J Cheminformatics* **2016,** *8*.
2. Djoumbou Feunang, Y.; Eisner, R.; Knox, C.; Chepelev, L.; Hastings, J.; Owen, G.; Fahy, E.; Steinbeck, C.; Subramanian, S.; Bolton, E.; Greiner, R.; Wishart, D. S., ClassyFire: automated chemical classification with a comprehensive, computable taxonomy. *J Cheminformatics* **2016,** *8* (1), 61.
3. Wishart, D. S.; Feunang, Y. D.; Marcu, A.; Guo, A. C.; Liang, K.; Vazquez-Fresno, R.; Sajed, T.; Johnson, D.; Li, C.; Karu, N.; Sayeeda, Z.; Lo, E.; Assempour, N.; Berjanskii, M.; Singhal, S.; Arndt, D.; Liang, Y.; Badran, H.; Grant, J.; Serra-Cayuela, A.; Liu, Y.; Mandal, R.; Neveu, V.; Pon, A.; Knox, C.; Wilson, M.; Manach, C.; Scalbert, A., HMDB 4.0: the human metabolome database for 2018. *Nucleic Acids Res* **2018,** *46* (D1), D608-D617.
4. *cxcalc*, 16.11; ChemAxon: 2017.
5. Dobson, C. M., Chemical space and biology. *Nature* **2004,** *432* (7019), 824-828.
6. Rosenberg, R. N.; Robinson, A. B.; Partridge, D., Urine Vapor Pattern for Olivopontocerebellar Degeneration. *Clin Biochem* **1975,** *8* (6), 365-368.
7. Dirren, H.; Robinson, A. B.; Pauling, L., Sex-related patterns in the profiles of human urinary amino acids. *Clin Chem* **1975,** *21* (13), 1970-5.
8. Robinson, A. B.; Dirren, H.; Sheets, A.; Miquel, J.; Lundgren, P. R., Quantitative Aging Pattern in Mouse Urine Vapor as Measured by Gas-Liquid-Chromatography. *Exp Gerontol* **1976,** *11* (1-2), 11-16.
9. Gates, S. C.; Sweeley, C. C.; Krivit, W.; Dewitt, D.; Blaisdell, B. E., Automated Metabolic Profiling of Organic-Acids in Human Urine .2. Analysis of Urine Samples from Healthy Adults, Sick Children, and Children with Neuroblastoma. *Clin Chem* **1978,** *24* (10), 1680-1689.
10. Politzer, I. R.; Githens, S.; Dowty, B. J.; Laseter, J. L., Gas chromatographic evaluation of the volatile constituents of lung, brain and liver tissues. *J Chromatogr Sci* **1975,** *13* (8), 378-9.
11. Nicholson, J. K.; Buckingham, M. J.; Sadler, P. J., High resolution 1H n.m.r. studies of vertebrate blood and plasma. *Biochem J* **1983,** *211* (3), 605-15.
12. Bales, J. R.; Higham, D. P.; Howe, I.; Nicholson, J. K.; Sadler, P. J., Use of High-Resolution Proton Nuclear Magnetic-Resonance Spectroscopy for Rapid Multi-Component Analysis of Urine. *Clin Chem* **1984,** *30* (3), 426-432.
13. Flint, H. J.; Porteous, D. J.; Kacser, H., Control of the Flux in the Arginine Pathway of Neurospora-Crassa - the Flux from Citrulline to Arginine. *Biochem J* **1980,** *190* (1), 1-15.
14. Middleton, R. J.; Kacser, H., Enzyme variation, metabolic flux and fitness: alcohol dehydrogenase in Drosophila melanogaster. *Genetics* **1983,** *105* (3), 633-50.
15. Rashed, M. S., Clinical applications of tandem mass spectrometry: ten years of diagnosis and screening for inherited metabolic diseases. *J Chromatogr B Biomed Sci Appl* **2001,** *758* (1), 27-48.
16. Clayton, P. T., Applications of mass spectrometry in the study of inborn errors of metabolism. *J Inherit Metab Dis* **2001,** *24* (2), 139-50.
17. Kuhara, T., Gas chromatographic-mass spectrometric urinary metabolome analysis to study mutations of inborn errors of metabolism. *Mass Spectrom Rev* **2005,** *24* (6), 814-27.
18. Castle, A. L.; Fiehn, O.; Kaddurah-Daouk, R.; Lindon, J. C., Metabolomics Standards Workshop and the development of international standards for reporting metabolomics experimental results. *Brief Bioinform* **2006,** *7* (2), 159-65.
19. Sumner, L. W.; Amberg, A.; Barrett, D.; Beale, M. H.; Beger, R.; Daykin, C. A.; Fan, T. W.; Fiehn, O.; Goodacre, R.; Griffin, J. L.; Hankemeier, T.; Hardy, N.; Harnly, J.; Higashi, R.; Kopka, J.; Lane, A. N.; Lindon, J. C.; Marriott, P.; Nicholls, A. W.; Reily, M. D.; Thaden, J. J.; Viant, M. R., Proposed minimum reporting standards for chemical analysis Chemical Analysis Working Group (CAWG) Metabolomics Standards Initiative (MSI). *Metabolomics* **2007,** *3* (3), 211-221.
20. Beisken, S.; Eiden, M.; Salek, R. M., Getting the right answers: understanding metabolomics challenges. *Expert Rev Mol Diagn* **2015,** *15* (1), 97-109.
21. Tulp, M.; Bohlin, L., Functional versus chemical diversity: is biodiversity important for drug discovery? *Trends Pharmacol Sci* **2002,** *23* (5), 225-231.
22. Fiehn, O., Metabolomics--the link between genotypes and phenotypes. *Plant Mol Biol* **2002,** *48* (1-2), 155-71.
23. Pence, H. E.; Williams, A., ChemSpider: An Online Chemical Information Resource. *J Chem Educ* **2010,** *87* (11), 1123-1124.

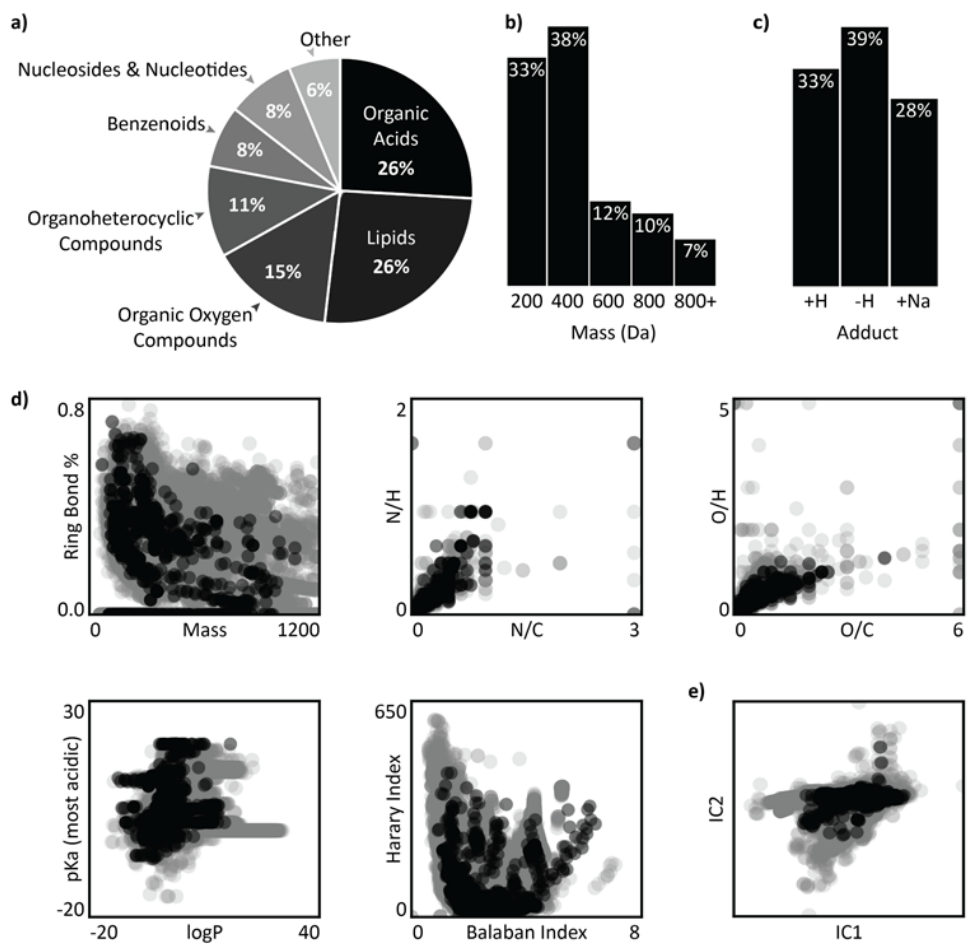

**Figure 1:** Validation set property distribution and chemical space coverage. (a) Superclass distribution of compounds, as determined by ClassyFire[2]. (b) Mass distribution with mass labels corresponding to (X-200, X]. (c) Adduct distribution. (d-e) Comparison of the validation set to the Human Metabolome Database (HMDB)[3], with black points corresponding to compounds found in the validation set and gray points corresponding to compounds found in the HMDB (v4.1, only those with masses 50-1200). (d) Distribution of predicted properties, with the ring bond percentage (number of bonds in rings divided by the total number of bonds), logP, pka, Balaban index, and Harary index calculated using *cxcalc*[4]. (e) Independent component analysis performed on the properties plotted in (d), with properties normalized to have a mean of 0 and standard deviation of 1.

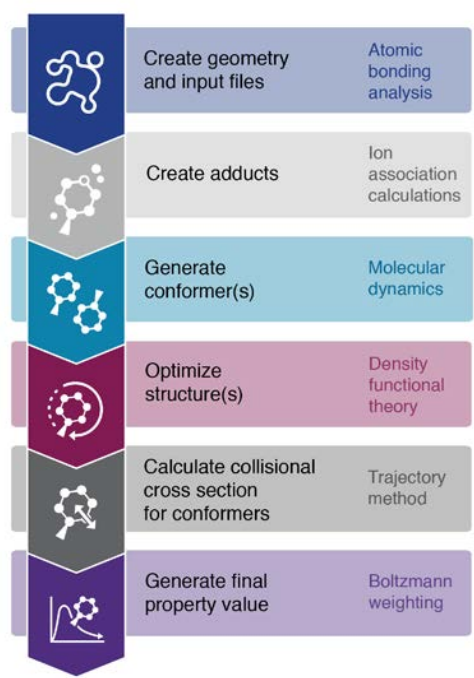

**Figure 2:** Schematic overview of ISiCLE module for CCS calculation. Major computational tasks are listed for the *Standard* method and, where appropriate, the associated method used. Tasks include preparation of input geometry from InChI, adduct formation, conformer generation by molecular dynamics, structure optimization by density functional theory, CCS calculation by the trajectory method, and, finally, final CCS prediction by Boltzmann weighting across conformers.

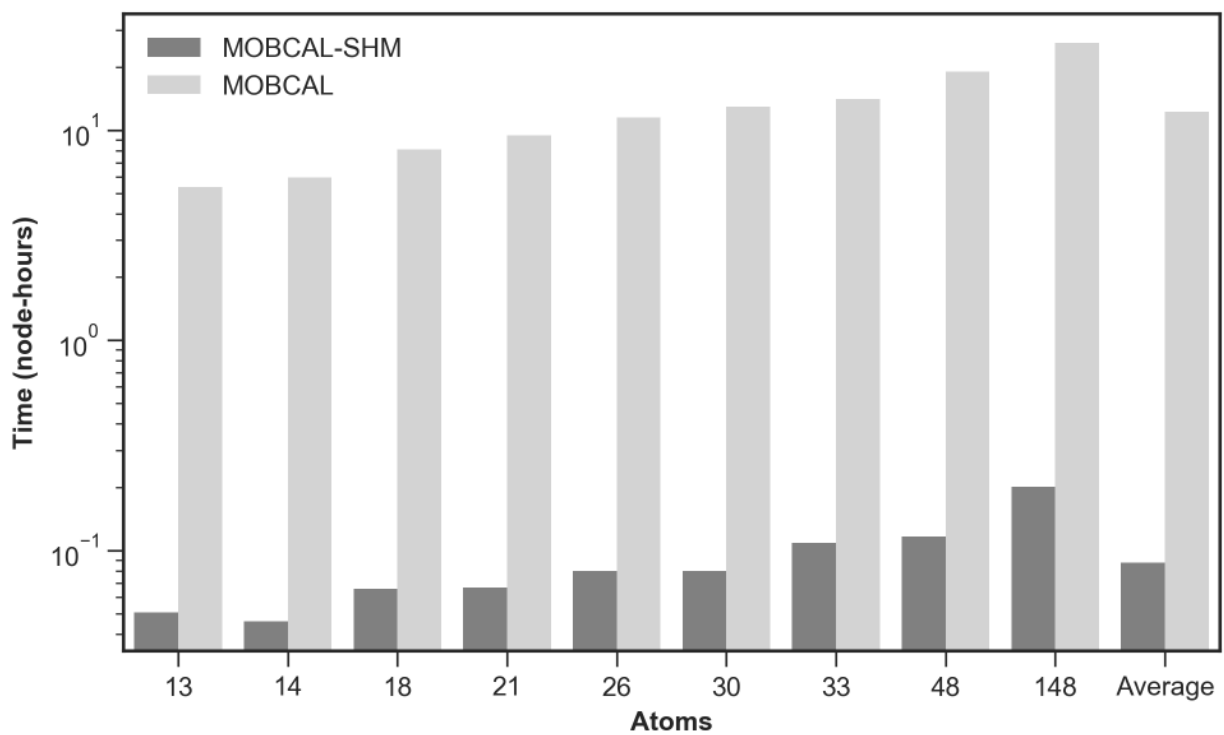

**Figure 3:** MOBCAL-SHM benchmark. Runtime comparison between the original MOBCAL implementation and the optimized, shared-memory version, MOBCAL-SHM. Times are shown for each atom size in the benchmark set, as well as the average. Improvements found in MOBCAL-SHM result in a 143-fold decrease in computation time without affecting prediction accuracy.

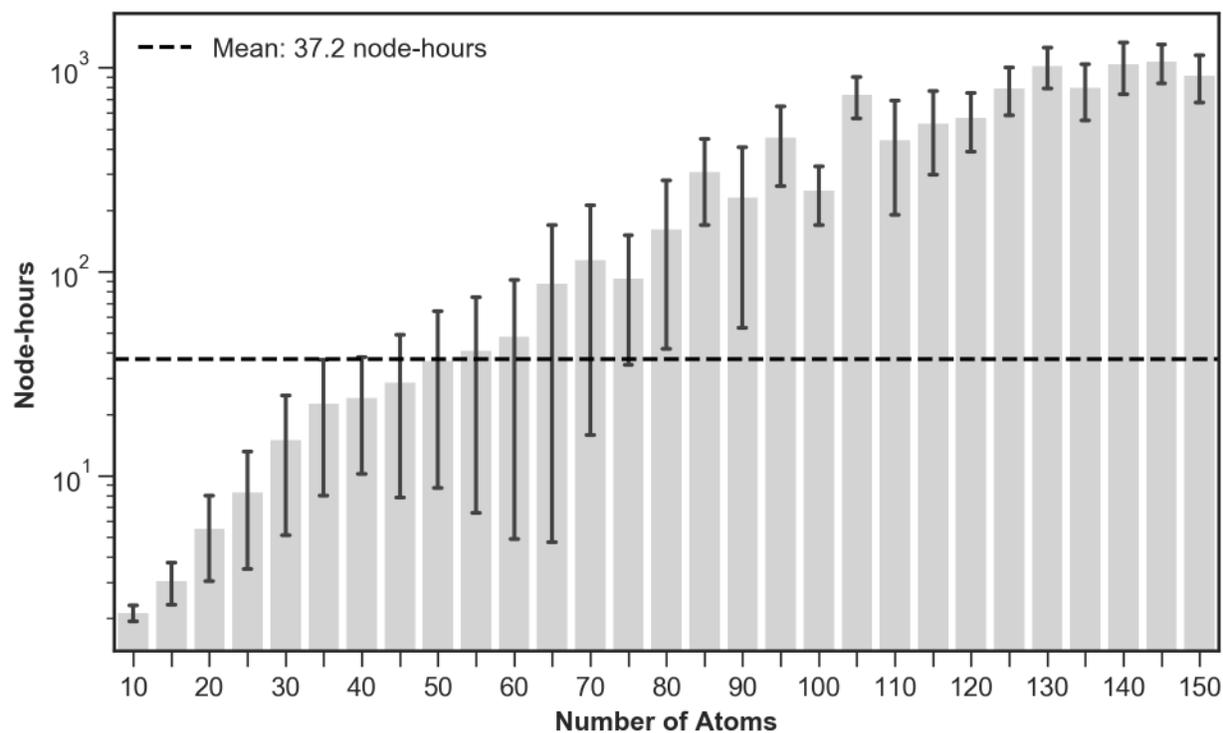

**Figure 4:** ISiCLE computational cost. Here we report total processing time per adduct ion in node-hours on a log scale. Adduct ions were grouped by number of atoms with a bin size of 5, where bin centers are reported on the x-axis. Error bars signify the standard deviation of observed times in each bin. Computational cost depends heavily on molecule size, but also varies within bins, particularly for mid-sized molecules (relative to this dataset).

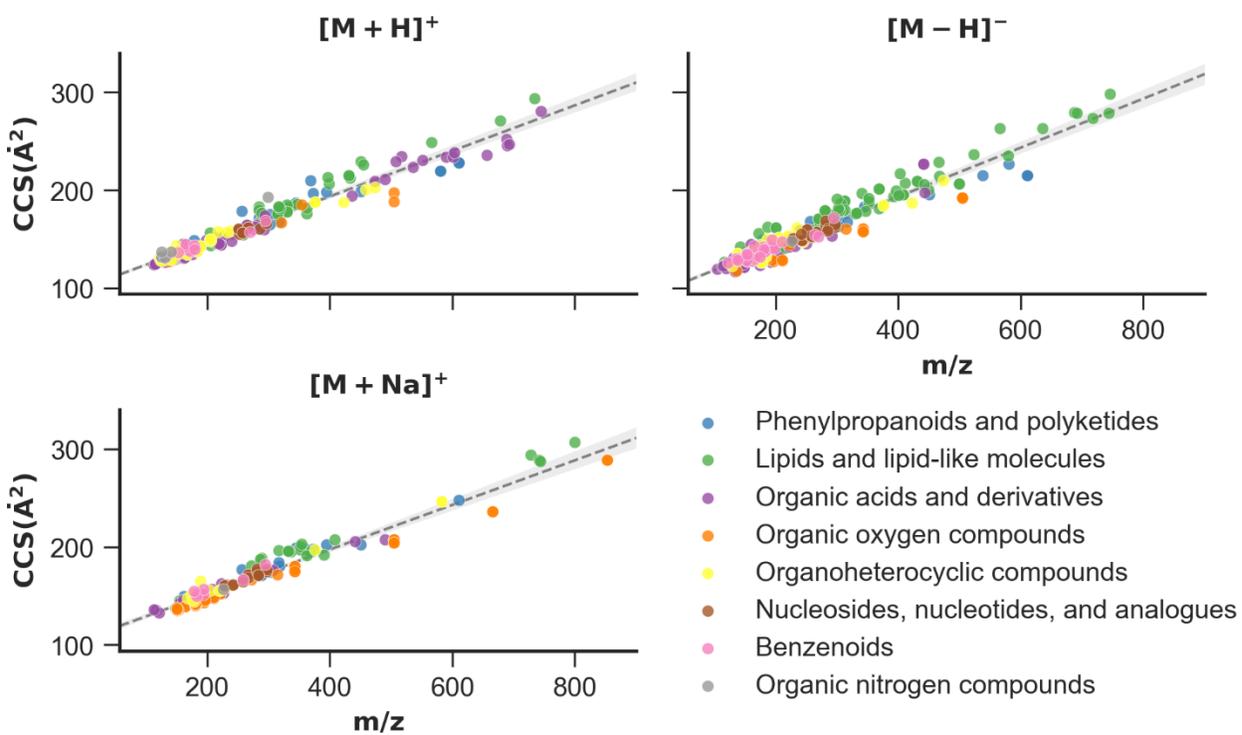

**Figure 5:** Calculated CCS versus *m/z*. Visual representation of CCS values calculated by ISiCLE *Standard* for the validation set, plotted against *m/z* by adduct ion, colored by chemical class as determined by ClassyFire[1].

**Table 1: Computational Burden of ISiCLE CCS Calculation Steps**

Each discrete task performed by the *Standard* method of ISiCLE is shown, as well as the software utilized, task scope (i.e. whether applied per molecule, adduct, or conformer), and computational burden. Computational burden is calculated based on average compute time for each task, adjusted based on task such that values are reported per adduct. Structure optimization by density NWChem represents the largest computational burden, followed by CCS calculation via MOBCAL-SHM and conformer generation via molecular dynamics. Remaining tasks did not demand appreciable computational cost.

| Task | Software | Per | Computational Burden |
|---|---|---|---|
| InChI Processing | ChemAxon, OpenBabel | Parent molecule | <1% |
| Create adducts | ChemAxon, OpenBabel | Adduct | <1% |
| Generate conformers | AMBER | Adduct | 1.2% |
| Optimize structure | NWChem | Conformer | 90.3% |
| CCS calculation | MOBCAL-SHM | Conformer | 7.4% |
| Boltzmann weighting | Python | Adduct | <1% |

**Table 2: Performance Comparison**

Mean absolute error (MAE) is shown for each method and dataset, where applied. The hierarchy of ISiCLE methods (*Lite, Standard, AIMD-based*) is captured, as well as ISiCLE's performance relative to similar methods (Paglia et al.[54], Bowers et al.[55]) and the machine learning-based MetCCS[40].

| Method | Validation Set | PAH Degradation Product (Application 1) | EPA ENTACT Mixtures (Application 2) | diCQA Isomers (Application 3) |
|---|---|---|---|---|
| ISiCLE (*Lite*) | 6.4% | 1.1% | 5.4% | 4.8% |
| ISiCLE (*Standard*) | 3.2% | -- | 3.1% | 2.6% |
| ISiCLE (*AIMD-based*) | -- | -- | -- | 0.8% |
| MetCCS | 3.3% | -- | -- | 6.4% |
| Paglia et al. | 5.3% | -- | -- | 3.1% |
| Bowers et al. | 5.2% | -- | -- | 3.7% |